\documentclass[10pt, twocolumn]{IEEEtran}

\usepackage{graphicx}

\usepackage{amsopn}
\usepackage{multirow}
\usepackage{amsxtra,amssymb}
\usepackage[english]{babel}
\ifCLASSINFOpdf
\else
\fi

\begin{document}
%
\title{A 3D Spatial Fluid Model for Wireless Networks}

%

\author{
\IEEEauthorblockN{
Jean-Marc Kelif$^1$,\thanks{$^1$Jean-Marc Kelif is with Orange Labs, Chatillon, France. Email: jeanmarc.kelif@orange.com}
Marceau Coupechoux$^2$\thanks{$^2$Marceau Coupechoux is with LTCI, Telecom ParisTech, France. Email: marceau.coupechoux@telecom-paristech.fr} 
}}

\maketitle

 \pagenumbering{gobble}

\begin{abstract}
In this article we develop a three dimensional (3D) analytical model of wireless networks. 
We establish an analytical expression of the SINR (Signal to Interference plus Noise Ratio) of user equipments (UE), 
by using a 3D fluid model approach of the network. 
This model enables to evaluate in a simple way the cumulative distribution function of the SINR, 
and therefore  the performance, the quality of service and the coverage of wireless networks, 
with a high accuracy. The use of this 3D wireless network model, instead of a standard two-dimensional one, in order to analyze wireless networks, is particularly interesting. 
Indeed, this 3D model enables to establish more accurate performance and quality of services results than a 2D one. 
\end{abstract}


%

\section{Introduction}

Most of analysis in wireless networks consider a 2D analytical model of network.
However since the needs are increasing in terms of QoS and performance, and because of the scarcity of  the frequency bandwidth, forecast of performance needs to be more and more accurate. This is the reason why more accurate analytical models are developed and in particular 3D models of wireless networks which should be closer to a real network.


A large literature has been developed about the modeling
of wireless networks \cite{Bon03} \cite{Vit95} \cite{Lagr05} \cite{KeA05}  \cite{KelCoEurasip10} \cite{GuK2000} \cite{LiP11} \cite{BB09} \cite{NBK07}, in the aim to analyze the capacity, throughput, coverage, and more generally performances for different types of wireless systems such as sensors, cellular and ad-hoc ones. 
These models of networks are based on i) stochastic geometry: the transmitters are distributed according to a spatial Poisson Process, ii) hexagonal pattern: the transmitting base stations constitute a regular infinite
hexagonal grid , iii) a 'fluid' approach: the interfering transmitters are replaced by a continuum. 
The hexagonal wireless network model is the most used one (\cite{Bon03}-\cite{Vit95}-\cite{Lagr05}). This model seems rather
'reasonable' for regular deployments of base stations. However, it is a two dimensional one: it does not take into account the height of antennas, and the propagation in the three dimensions. Moreover, from an analytical point of view, it is
intractable. Therefore, extensive computations are needed to establish performance.
Among the techniques developed to perform such computations, Monte Carlo simulations are widely used in
conjunction with this model \cite{Gil91,Ela05} or numerical computations
in hexagonal networks \cite{Vit94,Vit95}.
More generally, most of the works focus on 2D models of wireless networks. 

Nevertheless, 3D models of wireless networks were developed and analyzed in term of capacity \cite{GuK2000} \cite{LiP11}.
Authors of \cite{AlGo14} present a 3D geometrical propagation  mathematical model. Based on this model, a 3D reference model for MIMO mobile to mobile fading channels is proposed. 
In \cite{AMo15}, authors propose an exact form of the coverage probability, in two cases: i) interferers form a 3D Poisson Point Process and 
ii) interferers form a 3D  Modified Matern Process.  
The results established are 
then compared with a 2D case. 
Authors of \cite{KAC15} develop a  three-dimensional  channel  model between radar and cellular base stations, in which the radar uses  a  two-dimensional antenna array and the base station uses a one-dimensional antenna array. Their approach allows them to evaluate the interference impact.
The 3D  wireless model  for sensor  networks developed in \cite{XuLyH15} allows the authors to develop an analysis where the coverage can be improved. 
The tilt angle is the corner stone of an algorithm which uses the 3D model they propose,


\underline{Our Contribution}: In this paper, we develop a three dimensional analytical model of wireless networks. We establish a closed form formula of the SINR. We show that the formula allows analyzing wireless networks with more accuracy than a 2D wireless approach. Comparisons of performance coverage and QoS results, established with this 3D spatial fluid model and with a classical 2D model, show the wide interest of using this approach instead of a 2D approach. 

The paper is organized as follows. In Section \ref{model}, we develop the 3D network model. 
In Section \ref{3DFluid}, the analytical expression of the SINR by using this 3D analytical fluid model is established. 
In Section \ref{valid3D}, the validation of this analytical 3D model is done by comparison with Monte Carlo simulations.  
Section \ref{conclusion} concludes the paper.

\section{System Model} \label{model}

We consider a  wireless network consisting of $S$ geographical sites, each one composed by 3 base stations. Each antenna covers a sectored cell. We focus our analysis on the downlink, in the context of an OFDMA based wireless network, with frequency reuse 1.
Let us consider:
\begin{itemize}

\item ~${\mathcal S}=\{1,\ldots,S\}$ the set of geographic sites, uniformly and regularly distributed over a two-dimensional plane. 

\item ~${\mathcal N}=\{1,\ldots,N\}$ the set of base stations, uniformly and regularly distributed over the two-dimensional plane. The base stations are equipped with directional antennas: $N$= 3 $S$. 

\item the antenna height, denoted $h$. 

\item $F$ sub-carriers $f\in\mathcal{F}=\{1,\ldots,F\}$ where we denote $W$ the bandwidth of each sub-carrier.

\item $P_{f}^{(j)}(u)$ the transmitted power assigned by the base station $j$ to sub-carrier~$f$ towards user $u$. 
\item $g_{f}^{(j)}(u)$ the propagation gain between transmitter~$j$ and user $u$ in sub-carrier~$f$.
\end{itemize}

We assume that time is divided into slots. Each slot consists in a given sequence of OFDMA symbols. As usual at network level, we assume that there is no Inter-Carrier Interference (ICI) so that there is no intra-cell interference. 

The total amount of power received by a UE $u$ connected to the base station $i$, on sub-carrier $f$ is given by the sum of: a useful signal $P_{f}^{(i)}(u) g_{f}^{(i)}(u)$, an interference power due to the other transmitters $\sum\limits_{j\in\mathcal{N},j\neq i}P_{f}^{(j)}(u) g_{f}^{j}(u)$ and thermal noise power $N_{th}$.

We consider the SINR $\gamma_{f}(u)$ defined by:
\begin{equation} \label{SINR}
\gamma_{f}(u)=\frac{P_{f}^{(i)}(u) g_{f}^{(i)}(u) }{
\sum\limits_{j\in\mathcal{N},j\neq i}P_{f}^{(j)}(u) g_{f}^{j}(u)  + N_{th}}
\end{equation}
as the criterion of radio quality. 

We investigate the quality of service and performance issues of a network composed of sites equipped with 3D directional transmitting antennas. The analyzed scenarios consider that all the subcarriers are allocated to UEs (full load scenario). Consequently, each sub-carrier $f$ of any base station is used and can interfere with the ones of other sites. 
All sub-carriers are independent, we can thus focus on a generic one and drop the index $f$.

\subsection{Expression of the SINR} \label{SINRexpression}
Let us consider the path-gain model $g(R) = KR^{-\eta} A$, where  $K$ is a constant, $R$ is the distance between a transmitter $t$ and a receiver $u$, and $\eta >$ 2 is the path-loss exponent. The parameter $A$ is 
the antenna gain (assuming that receivers have a 0 dBi antenna gain).

\begin{figure}[htbp]
\centering
\includegraphics[scale=0.60]{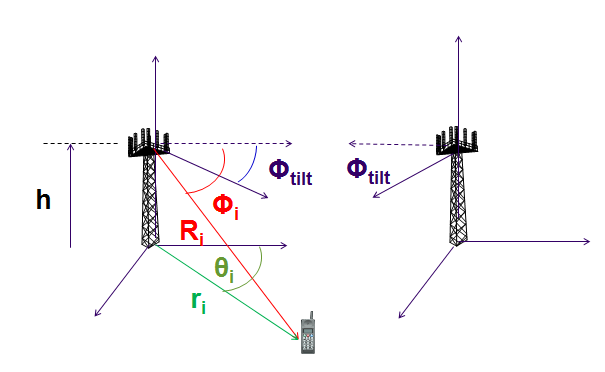}
\caption{\footnotesize User equipment located at $(r_i, \theta_i)$. It receives a useful power from antenna $i$ and interference power from antenna $j$. }
\label{antennes3D-2}
\end{figure}

Therefore,  for a user $u$ located at distance $R_i$ from its serving base station $i$, the expression (\ref{SINR}) of the SINR can be expressed, for each sub-carrier (dropping the index $f$): 

\begin{equation} \label{SINRdirect}
\gamma(R_i,\theta_i, \phi_i)=\frac{G_0 P KR_i^{-\eta}A(\theta_i,\phi_i) }{
G_0\sum\limits_{j\in\mathcal{N},j\neq i}P KR_j^{-\eta}A(\theta_j,\phi_j)  + N_{th}},
\end{equation}
where:
\begin{itemize}

\item $P$ is the transmitted power, 

\item $A(\theta_i, \phi_i)$ is the pattern of the 3D transmitting antenna of the base station $i$, and $G_0$ is the  maximum antenna gain.


\item $\theta_j$ is the horizontal angle between the UE and the principal direction
of the antenna $j$,

\item $\phi_j$ is the vertical angle between the UE and the antenna $j$ (see Fig.~\ref{antennes3D-2}),

\item $R_i = \sqrt{r_i^2+h^2}$, where $r_i$ represents the projection of $R_i$ on the ground.
\end{itemize}

The gain $G(\theta, \phi)$ of an antenna in a direction $(\theta, \phi)$ is defined as the ratio between the power radiated in that direction and the power that radiates an isotropic antenna without losses. This property characterizes the ability of an antenna to focus the radiated power in one direction.
The parameter $G_0$ (\ref{SINRdirect}) is particularly important for a beamforming analysis. Let notice that it is determined by considering that the power, which would be transmitted in all directions for a non directive antenna (with a solid angle of $4\pi$), is transmitted in a solid angle given by the horizontal and the vertical apertures of the antenna. In the ideal case where the antenna emits in a cone defined by $0 \leq \theta \leq \theta_{3dB}$ and $0 \leq \phi \leq \phi_{3dB}$, the gain is given by $\frac{4 \pi}{\int \int A(\theta, \phi) sin\theta d\theta d \phi}$. 


\subsection{BS Antenna Pattern} \label{sub:bsantennapattern}
In our analysis, we conform to the model of \cite{ITUR2009} for the antenna pattern (gain, side-lobe level). 
The antenna pattern which is applied to our scheme, is computed as: 
\begin{eqnarray} \label{eq:hv_pattern}
		A_{dB}(\theta, \phi) = -\min \left [ -(A_{h_{dB}}(\theta)+A_{v_{dB}}(\phi)), A_m \right ],
\end{eqnarray}
where $A_h(\theta)$ and $A_v(\phi)$ correspond respectively to the horizontal and the vertical antenna patterns.

The horizontal antenna pattern used for each base station is given by:
\begin{eqnarray} \label{eq:h_pattern}
		A_{h_{dB}}(\theta)= -\min \left[ 12 \left( \frac{\theta}{\theta_{3dB}} \right)^{2}, A_m \right ],
\end{eqnarray}
\noindent where:
\begin{itemize}
	\item $\theta_{3dB}$ is the half-power beamwidth (3 dB beamwidth);
	\item $A_m$ is the maximum attenuation.
\end{itemize}

The vertical antenna direction is given by:
\begin{eqnarray} \label{eq:v_pattern}
		A_{v_{dB}}(\phi)= -\min \left [ 12 \left( \frac{\phi-\phi_{tilt}}{\phi_{3dB}} \right)^{2}, A_m \right ],
\end{eqnarray}
\noindent where:
\begin{itemize}
	\item $\phi_{tilt}$ is the downtilt angle;
	\item $\phi_{3dB}$ is the 3 dB beamwidth.
\end{itemize}

\subsubsection{Antenna Pattern in the Network} \label{antennapatternnetwork}
Each site is constituted by 3 antennas (3 sectors). Therefore, for any site $s$ of the network, we have :
\begin{eqnarray}
	\left\{
		\begin{array}{ll}
				A_h(\theta_s^2) = A_h(\theta_s^1 + 2\frac{\pi}{3}) \\
				A_h(\theta_s^3) = A_h(\theta_s^1 - 2\frac{\pi}{3}) \\
				A_v(\phi_s^1) = A_v(\phi_s^2) = A_v(\phi_s^3),
		\end{array}
	\right.
\end{eqnarray}
where $\theta_s^a$ and $\phi_s^a$ represent the angles relative to the antenna $a \in \{1,2,3\}$ for the site $s$.
For the sake of simplicity, in expression (\ref{SINRdirect}) we do a sum on the base stations (not on the sites) and denote $\theta_j$ and $\phi_j$ the angles relative to the antenna~$j$.

\subsubsection{Vertical Antenna Gain in the Network} \label{verticalantennagain}
For a UE at the distance $r_j$ from the antenna $j$, the vertical angle can be expressed as:
\begin{equation} \label{phij}
\phi_j = \arctan \left(\frac{h}{r_j}\right).
\end{equation}
For interfering antennas, it can be noticed that since $r_j \gg h$, we have $\phi_j = \arctan \left(\frac{h}{r_j}\right) \rightarrow 0$, and  $\left( \frac{\phi_j-\phi_{tilt}}{\phi_{3dB}} \right)^{2} \rightarrow  \left( \frac{\phi_{tilt}}{\phi_{3dB}} \right)^{2}$.
Therefore the vertical antenna pattern (\ref{eq:v_pattern}) can be written as: 
\begin{eqnarray} \label{eq:v_pattern2}
A_{v_{dB}}(\phi)&=& -\min \left [ 12 \left( \frac{\phi-\phi_{tilt}}{\phi_{3dB}} \right)^{2}, A_m \right ] \nonumber \\
&\approx& -\min \left [ 12 \left( \frac{\phi_{tilt}}{\phi_{3dB}} \right)^{2}, A_m \right ] \nonumber \\
&=& G_{v_{dB}},
\end{eqnarray} 
where $G_{v_{dB}}= -\min \left [ 12 \left( \frac{\phi_{tilt}}{\phi_{3dB}} \right)^{2}, A_m \right ]$ (i.e. a constant).
And the antenna gain can be expressed as:

\begin{eqnarray} \label{eq:hv_pattern2}
A_{dB}(\theta, \phi) &=& -\min \left [ -A_{h_{dB}}(\theta)-A_{v_{dB}}(\phi)), A_m \right ] \nonumber \\
&=& -\min \left [ -A_{h_{dB}}(\theta) - G_{v_{dB}}  , A_m  \right ] \nonumber \\
&=& -\min \left [ -A_{h_{dB}}(\theta), A_m + G_{v_{dB}} \right ] +  G_{v_{dB}} \nonumber \\
&=& B_{dB}(\theta)+  G_{v_{dB}},
\end{eqnarray} 
where $B_{dB}(\theta) = -\min \left [ -A_{h_{dB}}(\theta), A_m + G_{v_{dB}} \right ]$.   

So we have:
\begin{eqnarray}
	\left\{
		\begin{array}{ll}
				B_{dB}(\theta) = -\min \left [ -A_{h_{dB}}(\theta), A_m + G_{v_{dB}} \right ] \\
				G_{v_{dB}} = - \min\left[12\left(\frac{\phi_{tilt}}{\phi_{3dB}}\right)^2,A_m\right]
		\end{array}
	\right.
\end{eqnarray}

Therefore, we establish that in this case, the vertical antenna gain only depends on the angle $\theta$.

\section{3D Wireless Network Spatial Fluid Model} \label{3DFluid}

The main assumption of the fluid network modeling \cite{KeA05} \cite{KelCoEurasip10} \cite{KeCoGo07}
consists in replacing a fixed finite number of eNBs by an
equivalent continuum of eNBs spatially distributed
in the network. 
Therefore, the transmitting power of the set of interfering 
eNBs of the network is considered as a continuum field all over the
network. 
Considering a density $\rho_{S}$ of sites $S$ and following the approach developed in  \cite{KeA05} \cite{KelCoEurasip10} \cite{KeCoGo07}, let us consider a UE located at $(R_i, \theta_i, \phi_i)$ in the area covered by the eNB $i$. 

Since each site is equipped by 3 antennas, we can express the denominator of (\ref{SINRdirect}) as:

\begin{eqnarray} \label{Interference}
I &=& \int 3 \times \rho_{S} K P R^{-\eta}A(\theta, \phi)t dt d\theta  \nonumber \\ 
&+& P KR_i^{-\eta}\sum_{a=2}^3 A(\theta_i^a, \phi_i^a)+ N_{th},
\end{eqnarray} 
where the integral represents the interference due to all the other sites of the network, and the discrete sum represents the interference due to the 2 antennas (eNB) co-localized with the eNB $i$. The index $a$ holds for these 2 antennas.


This can be further written as:

\begin{eqnarray}  
I &=& \int P \rho_{S} K (t^2+h^2)^{-\frac{\eta}{2}} t dt \times 3 \int A (\theta, \phi)d\theta
\nonumber \\
&+& P K (r_i^2+h^2)^{-\frac{\eta}{2}} \sum_{a=2}^3 A(\theta_i^a, \phi_i^a)+ N_{th}
\end{eqnarray}  
Since for the other eNB of the network, the distance $ r >> h $, we have $(t^2+h^2)^{-\frac{\eta}{2}} = t^{-\eta} (1+h^2/t^2)^{-\frac{\eta}{2}} \approx t^{-\eta}$, and the interference can be approximated by using (\ref{eq:hv_pattern2}): 

\begin{eqnarray}  \label{Interference2}
I &=& \int P \rho_{S} K t^{-\eta} t dt G_v \times 3 \int_0^{2 \pi} B (\theta) d\theta 
\nonumber \\
&+& P K (r_i^2+h^2)^{-\frac{\eta}{2}} \sum_{a=2}^3 A(\theta_i^a, \phi_i^a)+ N_{th}
\end{eqnarray} 

\subsection{3D Analytical Fluid SINR Expression} \label{3DFluid2}

The approach developed in \cite{KeCoGo07} \cite{KeCoGo10} allows to express $\int P \rho_{S} Kr^{-\eta} t dt$  as $\frac{\rho_{S} P K(2R_c-r_i)^{2-\eta}}{\eta-2}$, where $2 R_c$ represents the intersite distance (ISD).
We refer the reader to \cite{KelCoEurasip10} \cite{KeCoGo07} \cite{KeCoGo10} for the detailed explanation and validation through Monte Carlo simulations. 
Therefore, (\ref{Interference2}) can be expressed as:
\begin{eqnarray}  \label{interferencefluid}
I &=& \frac{3 G_v P K (2R_c-r_i)^{2-\eta}}{\eta-2} \rho_{S} \int_0^{2 \pi} B (\theta) d\theta  \nonumber \\ 
&+& P K (r_i^2+h^2)^{-\frac{\eta}{2}} \sum_{a=2}^3 A(\theta_i^a, \phi_i^a)+ N_{th}
\end{eqnarray}  

For a UE located at $(r,\theta,\phi)$ (dropping the index $i$), relatively to its serving eNB, the inverse of the SINR (\ref{SINRdirect}) is finally given by the expression : 
\begin{eqnarray}  \label{SINRfluid2}
\frac{1}{\gamma(r,\theta, \phi)}&=& \frac{3 G_v\rho_{S} (2R_c-r)^{2-\eta}}{(\eta-2)(r^2 +h^2)^{-\eta/2}} \frac{\int_0^{2 \pi} B (\theta) d\theta}{A(\theta, \phi)} \nonumber \\ 
&+& \frac{\sum_{a=2}^3 A(\theta^a, \phi^a)}{A(\theta, \phi)} \nonumber 
\\  
&+& \frac{N_{th}}{G_0 P K (r^2+h^2)^{-\eta/2} A(\theta, \phi)},
\end{eqnarray}  
where the index $a$ holds for the 2 antennas (eNB) co-localized with the serving antenna (eNB). 

%



%
%

\subsection{Interest of the SINR Analytical Formula}
The closed form formula (\ref{SINRfluid2}) allows the calculation of the SINR in an easy way. First of all, it only depends on the distance of a UE to its serving eNB (not on the distances to each eNB (SINR expression (\ref{SINRdirect})).
This formula highlights the network characteristic parameters which have an impact on the SINR (path loss parameter, inter-site distance, antenna gain). 
A simple numerical calculation is needed, due to the tractability of that formula.

\subsection{Throughput Calculation}
The SINR allows calculating the  maximum theoretical achievable throughput $D_u$ of a UE $u$, by using Shannon expression. 
For a bandwidth $W$, it can be written:
\begin{equation} \label{Dugamma}
D_u = W \log_2(1+\gamma_u)
\end{equation}

\textbf{Remarks:} In the case of realistic wireless network systems, it can be noticed that the mapping between the SINR and the achievable throughput are established by the mean of \textit{level curves}.\\

\section{Validation of the analytical formula}\label{valid3D}

The validation of the analytical formula (\ref{SINRfluid2}) consists in the comparison of the results established by this formula, to the ones established by Monte Carlo simulations. 

\subsection{Assumptions} \label{assumptionsantanna2D}

Let us consider:
\begin{itemize}
\item A hexagonal network composed of sectored sites;
\item Three base stations per site;
\item The 2D model: the antenna gain of a transmitting base station is given in dB by:
\begin{equation} \label{GTheta}
G_T(\theta) = - \min\left[12\left(\frac{\theta}{\theta_{3dB}}\right)^2, A_m \right],
\end{equation}
where $\theta_{3dB}$ = $70^\circ$ and $A_m =21$ dB;
\item The 3D model: the antenna gain of a transmitting base station is given by expressions (\ref{eq:hv_pattern}) (\ref{eq:h_pattern}) (\ref{eq:v_pattern});
\item Analyzed scenarios corresponding to realistic situations in a network:
\begin{itemize}
\item Urban environment: Inter Site Distance ISD = 200m, 500m and 750m,
\item Antennas tilts: $20^\circ$, $30^\circ$, $40^\circ$.
\end{itemize}

\end{itemize}

\subsection{Simulations vs 3D Analytical Model}
\begin{figure}[htbp!]
\centering
\includegraphics[scale=0.5]{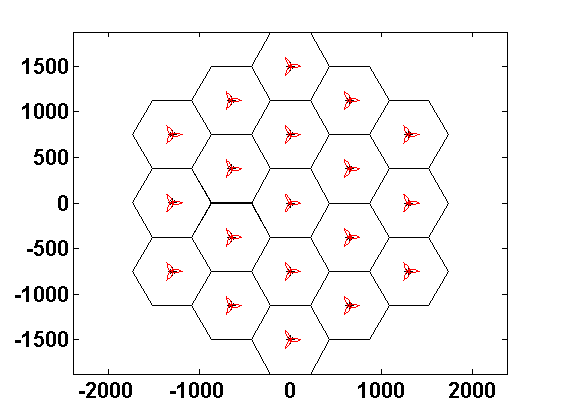}
\caption{\footnotesize Hexagonal network: location of the 3 sectors base stations in the plan. The $X$ and $Y$ axes represent the coordinates, in meters. The intersite distance in this example is 750 m.}
\label{hexagonaljpg}
\end{figure}

User equipments are randomly distributed in a cell of a 2D hexagonal based network (Fig.~\ref{hexagonaljpg}). This hexagonal network is equipped by antennas which have a given height (30m and 50m in our analysis), in the third dimension. Monte Carlo simulations are done to calculate the SINR for each UE. We focus our analysis on a typical hexagonal site. The cumulative distribution function (CDF) of the SINR can be established by using these simulations. These curves are compared to the ones established by using the analytical formula (\ref{SINRfluid2}) to calculate the SINR values. Moreover, the SINR values established by the two ways are drawn on figures representing a site with three antennas.

We present two types of comparisons. 
We first establish the CDF of the SINR. 
Indeed, the CDF of SINR provides a lot of information about the network characteristics: the coverage and the outage probability, the performance distribution, and the
quality of service that can be reached by the system. As an example, figure 3 shows that for an outage probability target of 10\%, the SINR reaches -8 dB, which corresponds to a given throughput.  
A second comparison, focused on the values of the SINR at each location of the cell, establishes a map of SINR over the cell. 

\subsection{Results of the Validation}
For the validation, we compare the two methods by considering realistic values of network parameters. An urban environment with realistic parameters of propagation is simulated \cite{ITUR2009}. Different tilts and apertures are considered.
The scenarios, summarized in Tab.~\ref{tab_scenariofigures}, show that the 3D fluid analytical model and the simulations provide very close values of SINR:

%

\begin{table}[!ht]
\caption{Scenarios and Figures}
\label{tab_scenariofigures}
\centering
\renewcommand{\arraystretch}{1.5}
	\begin{tabular}{|c|c|c|c|c|c|c| }
		\hline
		\textbf{Scenario} & \textbf{$\phi_{tilt}^{(\circ)}$}  & $\phi_{3dB}^{(\circ)}$  & $\theta_{3dB}^{(\circ)}$ & ISD (m) & h (m) & \textbf{Figures} \\ \hline
		Scenario 1 & 30 & 10 & 10 & 500 & 50 &3-4  \\ \hline
		Scenario 2 & 30 & 10 & 20 & 750 & 30 & 5-6-7  \\ \hline
		Scenario 3 & 20 & 10 & 10 & 750 & 30 & 8-9 \\    \hline
		Scenario 4 & 20 & 10 & 40 & 750 & 50 & 10-11 \\ \hline
		Scenario 5 & 40 & 30 & 20 & 750 & 30 & 12-13  \\ \hline
		Scenario 6 & 40 & 10 & 20 & 200 & 50 & 14-15  \\ \hline

		\end{tabular}
	\end{table}


\subsubsection{CDF of SINR}
The figures of scenario 1 (Fig.~3), scenario 2 (Fig.~5 and 6), scenario 3 (Fig.~8), scenario 4 (Fig.~10), scenario 5 (Fig.~12) and scenario 6 (Fig.~14) show that the analytical model (blue curves) and the simulations (red curves) provide very close CDF of SINR curves.

\subsubsection{Map of SINR}
The figures of scenario 1 (Fig.~4), scenario 2 (Fig.~7), scenario 3 (Fig.~9), scenario 4 (Fig.~11), scenario 5 (Fig.~13) and scenario 6 (Fig.~15) represent the values of SINR in each location of a cell, where the $X$ and $Y$ axes represent the coordinates (in meters). These figures show that the analytical model (right side) and the simulations (left side) provide very close maps of SINR.

 
\begin{figure}[htbp!]
\centering
\includegraphics[scale=0.3]{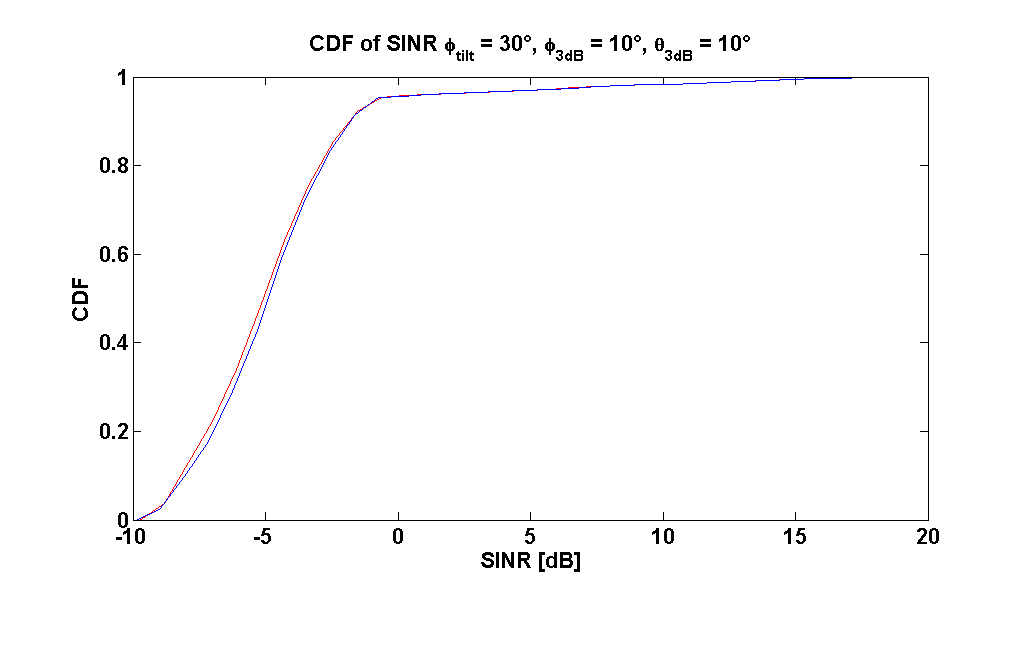}
\caption{\footnotesize Comparison of CDF of SINR for $\phi_{tilt} = 30^\circ$, a vertical  aperture $\phi_{3dB} = 10^\circ$  and an horizontal aperture $\theta_{3dB} = 10^\circ$.}
\label{tilt30phi3dB10theta3dB10}
\end{figure}

%

%
%
%

\begin{figure}[htbp!]
\centering
\includegraphics[scale=0.5]{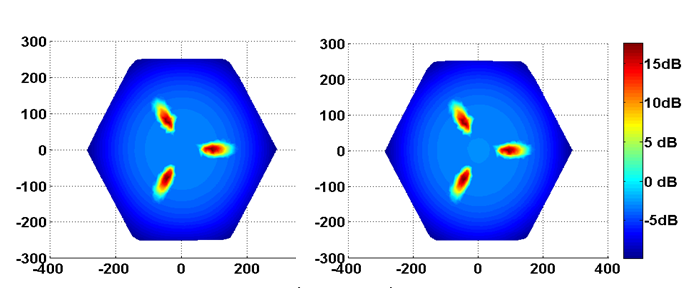}
\caption{\footnotesize Simulation (left) and Analytical (right) Map of the SINR for $\phi_{tilt} = 30^\circ$, a vertical  aperture $\phi_{3dB} = 10^\circ$  and an horizontal aperture $\theta_{3dB} = 10^\circ$.}
\label{AnalyticMaptilt30phi3dB10theta3dB10}
\end{figure}



\begin{figure}[htbp!]
\centering
\includegraphics[scale=0.3]{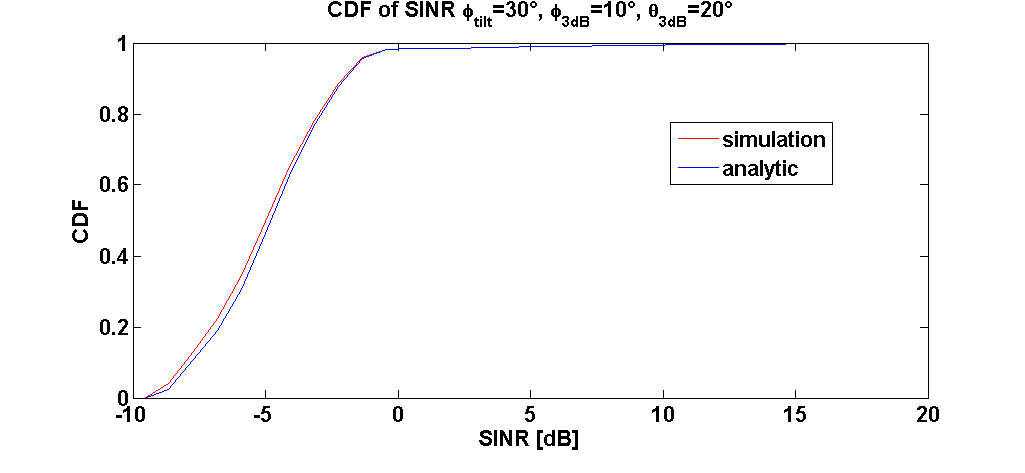}
\caption{\footnotesize Comparison of CDF of SINR for $\phi_{tilt} = 30^\circ$, a vertical  aperture $\phi_{3dB} = 10^\circ$  and an horizontal aperture  $\theta_{3dB} =20^\circ$.}
\label{tilt30phi3dB10theta3dB20}
\end{figure}

\begin{figure}[htbp!]
\centering
\includegraphics[scale=0.3]{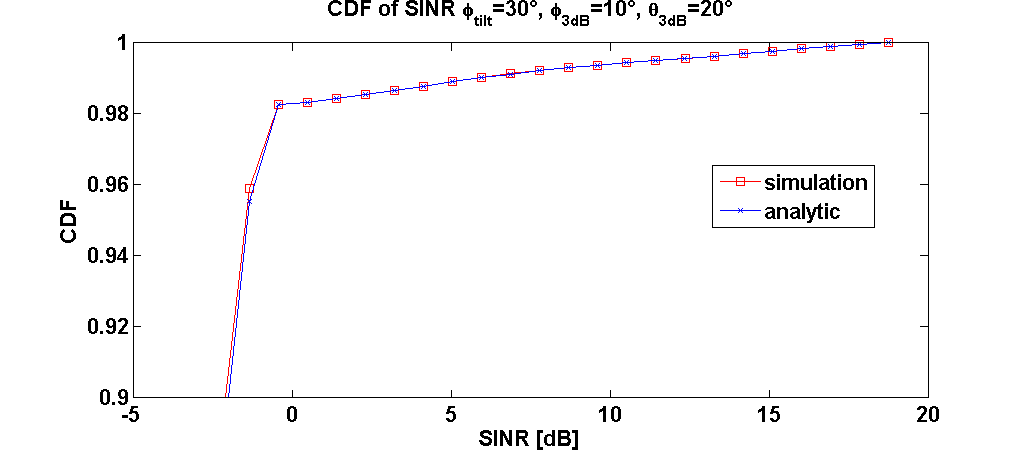}
\caption{\footnotesize Zoom on the upper part of the CDF (Fig \ref{tilt30phi3dB10theta3dB20}) where $\phi_{tilt} = 30^\circ$, $\phi_{3dB} = 10^\circ$  and $\theta_{3dB} = 20^\circ$.}
\label{Focustilt30phi3dB10theta3dB20}
\end{figure}

%
%
%

\begin{figure}[htbp!]
\centering
\includegraphics[scale=0.5]{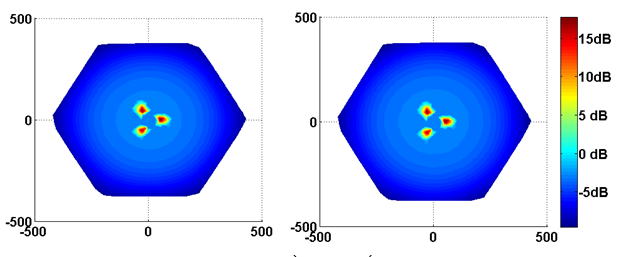}
\caption{\footnotesize Simulation (left) and Analytical (right) Map of the SINR for $\phi_{tilt} = 30^\circ$, a vertical  aperture $\phi_{3dB} = 10^\circ$  and an horizontal aperture $\theta_{3dB} = 20^\circ$.}
\label{AnalyticMaptilt30phi3dB10theta3dB20}
\end{figure}



\begin{figure}[htbp!]
\centering
\includegraphics[scale=0.3]{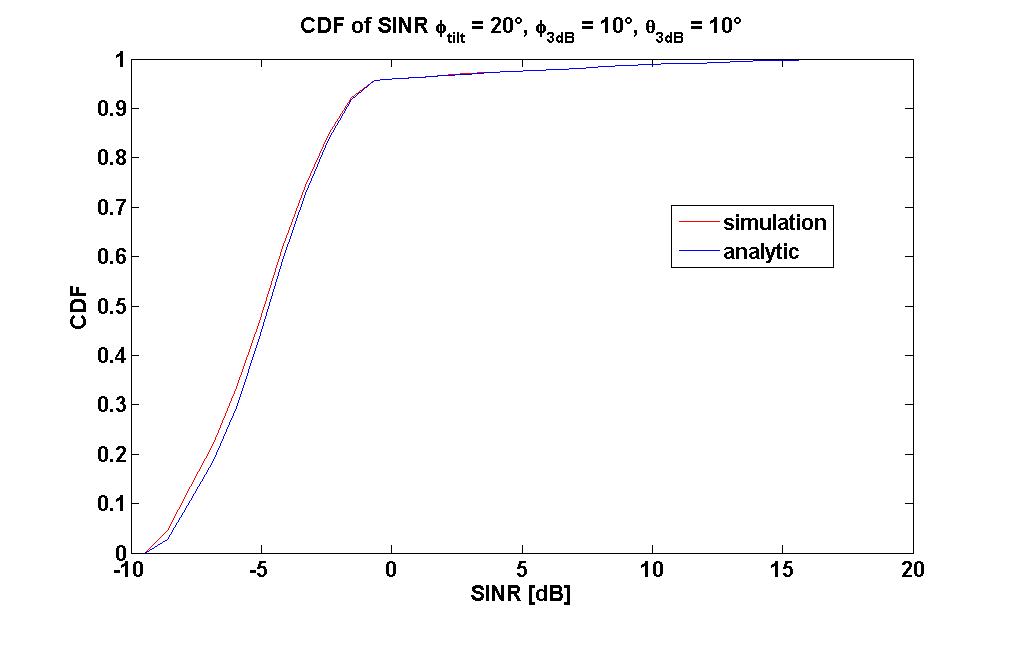}
\caption{\footnotesize Comparison of CDF of SINR for $\phi_{tilt} = 20^\circ$, a vertical  aperture $\phi_{3dB} = 10^\circ$  and an horizontal aperture $\theta_{3dB} = 10^\circ$.}
\label{tilt20phi3dB10theta3dB10}
\end{figure}

%
%
%

\begin{figure}[htbp!]
\centering
\includegraphics[scale=0.5]{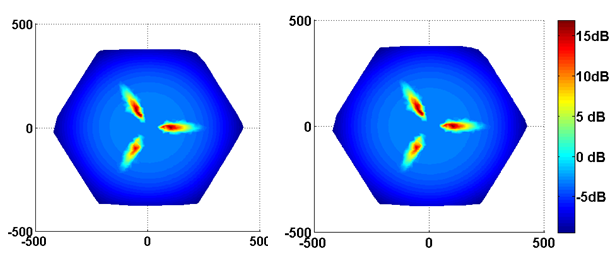}
\caption{\footnotesize Simulation (left) and  Analytical  (right) Map of the SINR for $\phi_{tilt} = 20^\circ$, a vertical  aperture $\phi_{3dB} = 10^\circ$  and an horizontal aperture $\theta_{3dB} = 10^\circ$.}
\label{SimuAnalyticMaptilt20phi3dB10theta3dB10}
\end{figure}



\begin{figure}[ht!]
\centering
\includegraphics[scale=0.3]{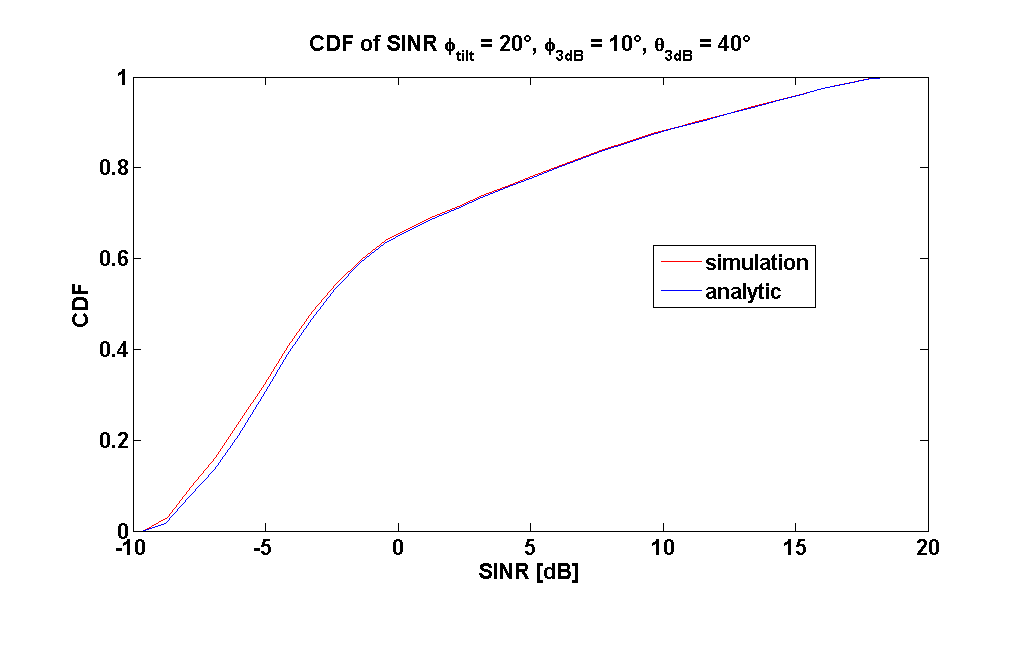}
\caption{\footnotesize Comparison of CDF of SINR for $\phi_{tilt} = 20^\circ$, a vertical  aperture $\phi_{3dB} = 10^\circ$  and an horizontal aperture $\theta_{3dB} = 40^\circ$.}
\label{tilt20phi3dB10theta3dB40}
\end{figure}

%
%
%
%
\begin{figure}[htbp!]
\centering
\includegraphics[scale=0.5]{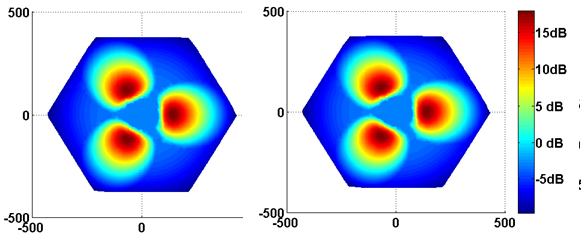}
\caption{\footnotesize Simulation (left) Analytical (right) Map of the SINR for $\phi_{tilt} = 20^\circ$, a vertical  aperture $\phi_{3dB} = 10^\circ$  and an horizontal aperture $\theta_{3dB} = 40^\circ$.}
\label{AnalyticMaptilt20phi3dB10theta3dB40}
\end{figure}



\begin{figure}[htbp!]
\centering
\includegraphics[scale=0.3]{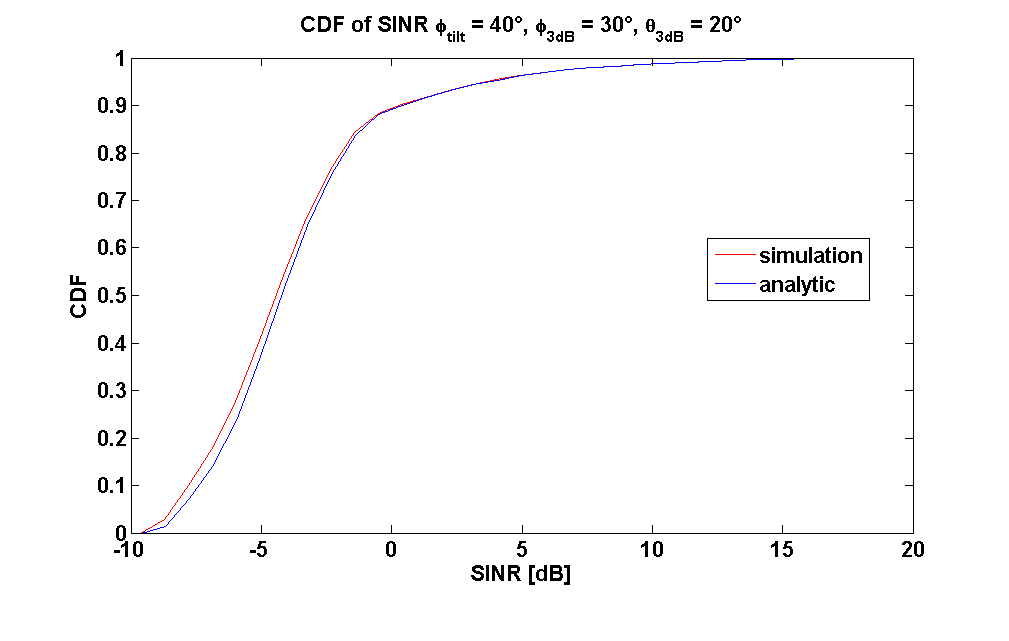}
\caption{\footnotesize Comparison of CDF of SINR for $\phi_{tilt} = 40^\circ$, a vertical  aperture $\phi_{3dB} = 30^\circ$  and an horizontal aperture $\theta_{3dB} = 20^\circ$.}
\label{tilt40phi3dB30theta3dB20ISD750H30}
\end{figure}

%
%
%
%

\begin{figure}[htbp!]
\centering
\includegraphics[scale=0.5]{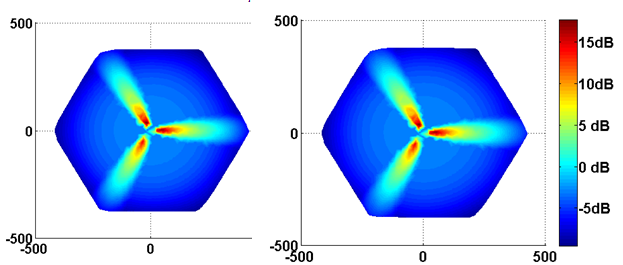}
\caption{\footnotesize Simulation (left) and Analytical (right) Map of the SINR for $\phi_{tilt} = 40^\circ$, a vertical  aperture $\phi_{3dB} = 30^\circ$  and an horizontal aperture $\theta_{3dB} = 20^\circ$.}
\label{AnalyticMaptilt40phi3dB30theta3dB20ISD750H30}
\end{figure}



\begin{figure}[htbp!]
\centering
\includegraphics[scale=0.3]{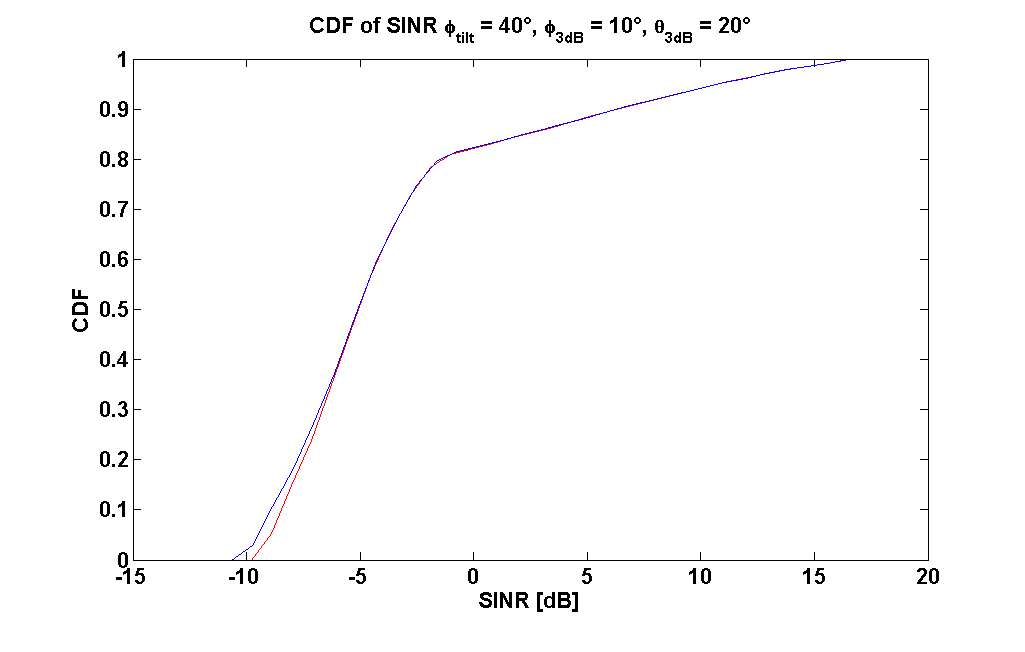}
\caption{\footnotesize Comparison of CDF of SINR for $\phi_{tilt} = 40^\circ$, a vertical  aperture $\phi_{3dB} = 10^\circ$  and an horizontal aperture $\theta_{3dB} = 20^\circ$.}
\label{tilt40phi3dB10theta3dB20ISD200H50}
\end{figure}

\begin{figure}[htbp!]
\centering
\includegraphics[scale=0.5]{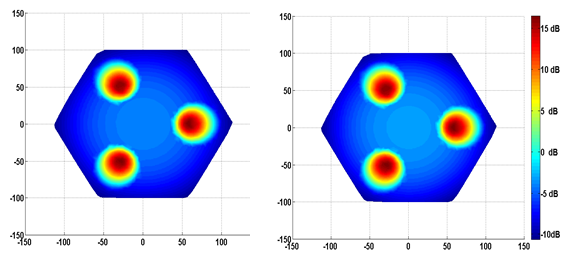}
\caption{\footnotesize Simulation (left) and Analytic (right) Map of the SINR for $\phi_{tilt} = 40^\circ$, a vertical  aperture $\phi_{3dB} = 10^\circ$  and an horizontal aperture $\theta_{3dB} = 20^\circ$.}
\label{SimuAnalyticMaptilt40phi3dB10theta3dB20ISD200H50}
\end{figure}

\subsection{Limitation of the 3D Fluid Model of Wireless Network} \label{LimitModel}


The aim of our analysis is to propose a model allowing to evaluate the performance, quality of service and coverage reachable in a cell whose standard antennas are replaced by 3D antennas, taking into account the height of the antenna and its tilt. Therefore the whole antenna energy is focused on the cell. This implies that the angle $\phi_{3dB}$ has to be lower than $\phi_{tilt}$, otherwise UEs belonging to other cells could be served by this antenna. The validation process was done according to this constraint.
However, the analytical closed-form formula (\ref{SINRfluid2}) allows to establish CDF of SINR very closed to simulated ones, for the different values of $\phi_{tilt}$, vertical apertures $\phi_{3dB}$ and horizontal apertures $\theta_{tilt}$, as soon as $\phi_{tilt}\geq \phi_{3dB}$.
Moreover, the SINR maps given by simulations and by the formula are also very closed.
Therefore, the formula is particularly well adapted for 3D wireless networks analysis.

\section{Conclusion} \label{conclusion}
We develop, and validate, a three dimensional analytical wireless network model. 
This model allows us to establish a closed form formula of the SINR reached by a UE at any location of a cell, for a 3D wireless network.
The validation of this model, by comparisons with Monte Carlo simulations results, shows that the two approaches establish very close results, in terms of CDF of SINR, and also in terms of SINR map of the cell. 
This model may be used in the aim to analyze wireless networks, in a simple way, with a higher accuracy than a classical 2D approach.


\nocite{*}
\end{document}